\documentclass[conference]{IEEEtran}
\IEEEoverridecommandlockouts
\IEEEpubid{\begin{minipage}{\textwidth}\ \\[12pt]
	978-1-7281-9818-7/20/\$31.00 \copyright 2020 IEEE
\end{minipage}}

\usepackage{cite}
\usepackage{amsmath,amssymb,amsfonts}
\usepackage{algorithmic}
\usepackage{graphicx}
\usepackage{textcomp}
\usepackage{xcolor}

\usepackage{graphicx} % \includegraphics with witdth etc.
\usepackage{tabularx}
\usepackage{booktabs}	% \toprule
\usepackage{array}
\usepackage{xfrac}	% \sfrac

\def\BibTeX{{\rm B\kern-.05em{\sc i\kern-.025em b}\kern-.08em
    T\kern-.1667em\lower.7ex\hbox{E}\kern-.125emX}}
\begin{document}

% ACIRS
\title{A Sensitivity Analysis Approach for Evaluating a Radar Simulation for Virtual Testing of Autonomous Driving Functions\\
}

\author{\IEEEauthorblockN{1\textsuperscript{st} Anthony Ngo}
	\IEEEauthorblockA{\textit{CC-Automated Driving} \\
		\textit{Robert Bosch GmbH}\\
		Stuttgart, Germany\\
		anthony.ngo@de.bosch.com}
	\and
	\IEEEauthorblockN{2\textsuperscript{nd} Max Paul Bauer}
	\IEEEauthorblockA{\textit{CC-Automated Driving} \\
		\textit{Robert Bosch GmbH}\\
		Stuttgart, Germany\\
		max.bauer@de.bosch.com}
	\and
	\IEEEauthorblockN{3\textsuperscript{rd} Michael Resch}
	\IEEEauthorblockA{\textit{High Performance Computing Center} \\
		\textit{University of Stuttgart}\\
		Stuttgart, Germany\\
		resch@hlrs.de}
}

\maketitle

\begin{abstract}
Simulation-based testing is a promising approach to significantly reduce the validation effort of automated driving functions. Realistic models of environment perception sensors such as camera, radar and lidar play a key role in this testing strategy. A generally accepted method to validate these sensor models does not yet exist. Particularly radar has traditionally been one of the most difficult sensors to model. Although promising as an alternative to real test drives, virtual tests are time-consuming due to the fact that they simulate the entire radar system in detail, using computation-intensive simulation techniques to approximate the propagation of electromagnetic waves.
In this paper, we introduce a sensitivity analysis approach for developing and evaluating a radar simulation, with the objective to identify the parameters with the greatest impact regarding the system under test.
A modular radar system simulation is presented and parameterized to conduct a sensitivity analysis in order to evaluate a spatial clustering algorithm as the system under test, while comparing the output from the radar model to real driving measurements to ensure a realistic model behavior. The presented approach is evaluated and it is demonstrated that with this approach results from different situations can be traced back to the contribution of the individual sub-modules of the radar simulation. 
\end{abstract}

\begin{IEEEkeywords}
	Virtual testing, autonomous driving, radar simulation, sensor modeling, automotive radar, environment perception, sensitivity analysis.
\end{IEEEkeywords}

\section{Introduction}  \label{sec1: Introduction}
Autonomous driving is currently one of the main trends in the automotive industry. Fully automated driving offers the greatest potential for minimizing the likelihood of accidents and optimizing traffic flow \cite{friedrich_effect_2016}. As the industry moves towards full automation, there is a growing need to develop not only advanced safety systems, but also the tools for their accurate analysis in order to validate the highly complex system \cite{koopman_challenges_2016}. A vast amount of driving kilometers are needed to statistically proof the safety of an autonomous vehicle \cite{wachenfeld_release_2016}. For instance to ensure that an autonomous vehicle can handle 95\% of the driven  kilometers safely, it would be necessary to drive a total of ten million kilometers. Even with all these kilometers it is not guaranteed that the right scenarios are considered and all critical situations tested. This indicates that with real driving tests alone a statistical homologation would not be economically feasible. The combination of real driving tests and simulation-based testing is a promising approach to significantly reduce the validation effort of autonomous driving functions \cite{jesenski_simulation-based_2019}.

Realistic models of environment perception sensors such as camera, radar and lidar play a key role in that testing strategy \cite{schaermann_validation_2017}. These sensor models have to be validated in order to permit any reliable predictions about the behavior of the real system through virtually testing of autonomous driving functions \cite{oberkampf_simulation_2019}.
While being considered as a key sensor for autonomous driving, radar has traditionally been one of the most difficult sensors to model \cite{wheeler_deep_2017}. There have been many different approaches to model a radar sensor system. Although a lot of radar effects are understood and can be modeled today, a high fidelity simulation faces challenges regarding the required computation time \cite{chipengo_antenna_2018}. This is due to the fact that radar exhibits numerous characteristics, including multipath reflections, interference, ambiguities, clutter, ghost objects, and attenuation \cite{skolnik_radar_2008}, which leads to exceedingly high demands on the computing power for a comprehensive and profound simulation. However, the question arises whether a detailed radar sensor model is required in all simulation scenarios. The problem to find the sufficient level of detail remains unsolved and the right trade-off between model realism and computation speed must be found.

Therefore, we introduce a sensitivity analysis approach for developing and validating a radar system simulation, with the aim to identify the radar sensor effects with the greatest impact considering a system under test (SUT). Focusing on the most important effects achieves a high radar model fidelity, while reducing the computation time needed. In this paper, we present a proof-of-concept implementation of the sensitivity analysis method analyzing a radar simulation for testing a clustering algorithm.

% TODO: double check at the end
The remainder of this paper is structured as follows. Section \ref{sec2: Related Work} gives a brief overview of existing radar simulation and sensor model evaluation approaches. Section \ref{sec3: Method} elaborates the proposed approach in detail. Based on this, Section \ref{sec4: Results} explains the conducted experiments and discuss their results. Finally, in Section \ref{sec5: Conclusion} a summary including a brief outlook on future works concludes this paper.

\section{Related Work} \label{sec2: Related Work}
The following section gives an overview of different radar sensor modeling approaches for testing autonomous driving functions and presents the state of the art in sensor model validation.

\subsection{Radar Sensor Modeling} \label{sec2.1: Radar Sensor Modeling}
There exist several approaches for modeling a radar sensor in the literature. What mainly differentiates them is the way they model the propagation of electromagnetic waves, which is a key part of any radar simulation. Although electromagnetic radiation is governed by Maxwell's equations, it is not feasible in general to have an analytical solution in a realistic propagation environment \cite{yun_ray_2015}.
There are different definitions for radar sensor model types. Usually these definitions relate to the level of detail with which a model embodies the real radar sensor.

% Numerical Methods
Time-domain electromagnetic simulation techniques such as the finite-difference time-domain method \cite{yee_numerical_1966}, the finite integration technique \cite{clemens_discrete_2001}, the finite element method \cite{jin_finite_2014}, and method of moments \cite{harrington_field_1968} are based on the spatiotemporal discretization of Maxwell's equations. They can be used for a detailed simulation of the electromagnetic phenomena observed in radar systems \cite{machida_rapid_2019}. Despite the fact that they are very precise in principle, assuming that the analysis space is large relative to the wavelength, numerical methods based on the discretization of differential or integral equations face the challenge of exorbitantly large memory needs and slow computational speed \cite{yun_ray_2015}. More specifically, if the simulation frequency is approximately 77 Ghz, which is within the scope of automotive radar systems, it is not feasible to simulate the space enclosing a whole vehicle as a consequence of a prohibitively long computation time \cite{owaki_hybrid_2019}.

% Ray Tracing Approach
A widely used method to overcome this problem is the ray tracing approach based on the geometric optics diffraction theory, in which radio waves are regarded as a bundle of rays \cite{keller_geometrical_1962}. % cite here some important RT paper?
Ray tracing enables to simulate various radar sensor effects like reflection, diffraction, radar ghost objects etc. \cite{yun_ray_2015}.
Although this approach requires less computation power than numerical methods, they are still computationally expensive, limiting their use in real-time applications like hardware in the loop setups \cite{holder_measurements_2018}. Apart from the limitation in execution speed, this approach demands a high level of detail for the simulation of the environment. Particularly geometry and material properties of all surrounding objects are a prerequisite for a high fidelity propagation model \cite{chipengo_antenna_2018}.

% data driven models
Data-driven sensor models strive to address this matter by learning from real sensor data, which inherently hold information about the observed environment. This method eliminates both the need to model the radar phenomena in detail and to have all the details about the surroundings. For this reason, these models are also called black-box models \cite{cao_modeling_2017}. Data-driven models can exhibit fundamental radar effects while remaining real-time capable \cite{wheeler_deep_2017}. Nevertheless, the drawback of this approach is that these type of models utterly depend on the available training data, which in most cases has been recorded on a test site for a simplified ground truth determination. As a result, the observed scenarios are usually restricted in terms of environment scenery and numbers. This makes it difficult to generalize to more complicated environments, because radar sensor data are prone to vary strongly depending on changes in the observed scenarios. 

% ideal sensor model
An ideal sensor model represents the simplest form of a radar sensor model, which considers merely the optical field of view without measurement errors, i.e. objects are detected any time they are within the sensor's measurement range. Radar sensor specific physical effects are not taken into account. These type of models are also called ground truth sensor models. Due to the their simplicity and fast computation time, they are particularly suitable for early testing of perception algorithms in either ideal conditions or under the assumption that sensor errors are neglectable \cite{holder_measurements_2018}.

\subsection{Radar Sensor Model Evaluation} \label{sec2.2: Radar Sensor Model Evaluation}
In order to derive conclusions about the real system behavior from synthetic sensor data, it is first necessary to determine the sufficient degree of realism, i.e. validating the radar model.

There exists no generally accepted evaluation criteria or requirements to compare the output from a radar sensor model to a real radar sensor \cite{holder_measurements_2018}. Therefore, several different approaches have been reported in the literature, which can be distinguished inter alia by the degree of abstraction of the radar model output: raw data level \cite{machida_rapid_2019}, \cite{cao_modeling_2017}; detection level \cite{martowicz_uncertainty_2019}; or object level \cite{roth_analysis_2011}, \cite{suhre_simulating_2018}. Accordingly, the abstract raw data level represents any level before a radar detection is generated, for example the received power amplitude, and is more complicated to compare due to the stochastic character of a radar sensor.

Besides the level of abstraction, two additional topics for sensor model validation can be differentiated. First, direct comparison of recorded sensor data with the direct model output. 

Second, comparison of the SUT results with the SUT being the subsequent stage after the sensor model in the system, which uses experimentally and synthetically generated radar sensor data as input. This can be for example a clustering or tracking algorithm.
Although direct comparison is necessary, it is not sufficient for sensor model validation \cite{schaermann_validation_2017}.
For example, despite that an ideal sensor model might lack accuracy in a direct comparison, the results from a subsequent algorithm can still show a great consensus. Consequently, sensor models in the scope of autonomous driving functions can not be treated as stand-alone applications, the system under test has to be considered \cite{rosenberger_towards_2019}.
Thus, a method that provides a quantitative comparison while considering the system under test is needed. Due to demanding requirements in terms of execution speed, furthermore, the sufficient degree of realism must be found.

\section{Method} \label{sec3: Method}
Considering the insights from Section \ref{sec2: Related Work}, the present section elaborates the proposed method, starting with a brief overview of the approach, followed by a more detailed explanation of the different components.

\subsection{Overview}
\begin{figure}[thpb]
	\centering
	\setlength{\fboxrule}{0pt}
	\framebox{\parbox{\linewidth}{
			\includegraphics[width=\linewidth] {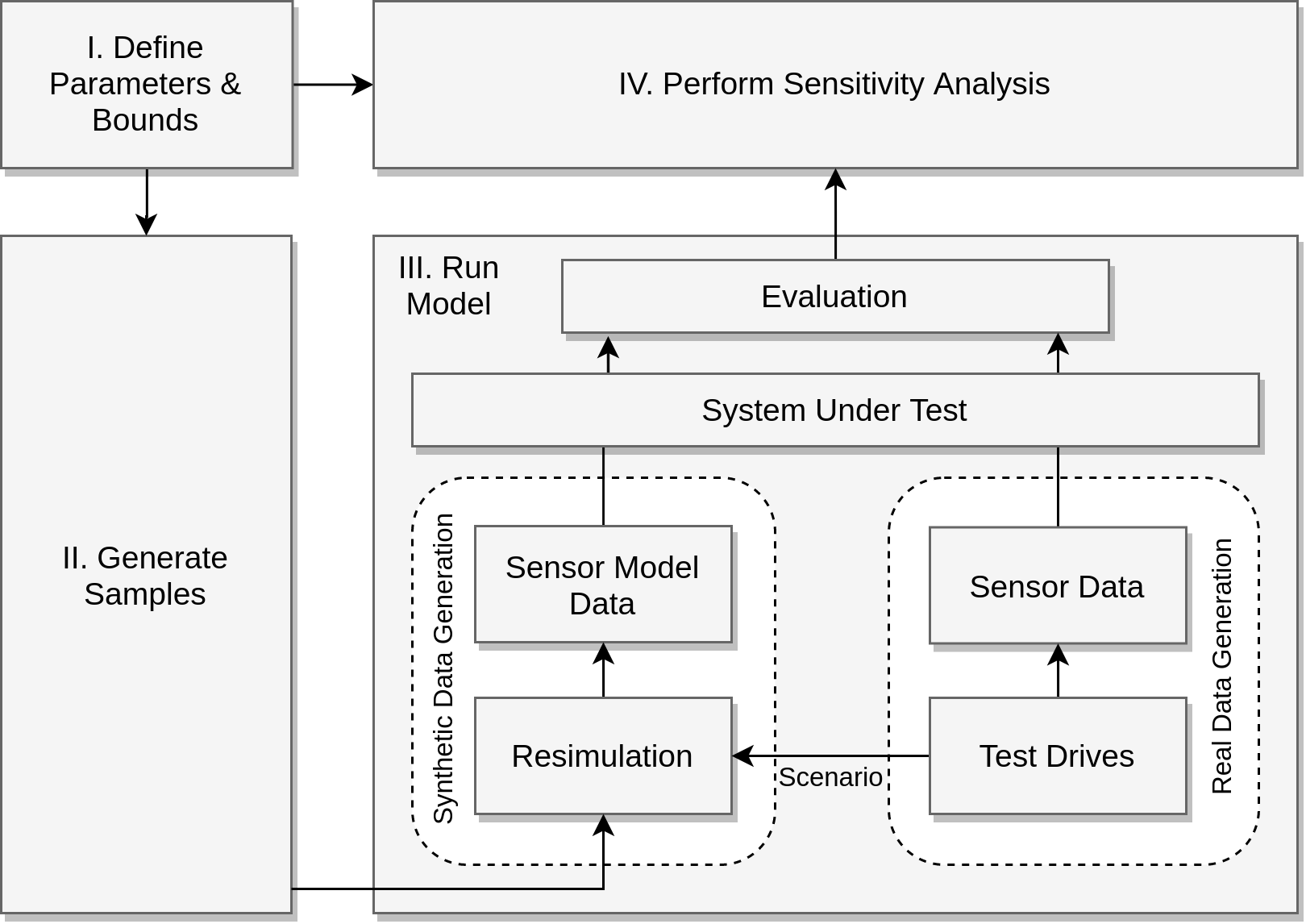}
	}}
	\caption{Sensitivity analysis approach for evaluating a radar sensor model}
	\label{fig: sensitivity analysis approach}
\end{figure}

The method introduced in this section focuses on the enhancement of the existing approaches by incorporating a quantitative evaluation of the implemented radar sensor effects. This is achieved by performing a sensitivity analysis to determine the impact of each effect on a system under test. The proposed approach consists of the several steps, which are illustrated in Fig. \ref{fig: sensitivity analysis approach}.

\begin{enumerate}
	\item At first, the sensor effects to be examined and their bounds are specified. A distinction is made between sensor-dependent and scenario-dependent parameters, while the latter is not in the scope of this particular work.
	\item Subsequently, the samples are generated, resulting in a matrix whose dimensions are defined by the number of samples as well as the number of parameters.
	\item In a third step, real test drives are carried out and are then simulated to capture the output from both sensor and sensor model respectively. Each are processed by the SUT and the results are evaluated based on a chosen metric.
	\item The actual senstivitiy analysis is conducted in the final step, which uses the defined parameters with their bounds and the evaluated model outputs as inputs, eventuating in the sensitivity indices. 
\end{enumerate}

\subsection{Real Data Generation}
The generation of real sensor data as a reference for comparison is an essential element for the sensor model evaluation. Therefore, it is necessary to determine the ground truth data, i.e. the correct position and orientation of surrounding objects from the perceived environment, which is not a trivial task. Due to the fact that the ground truth data serve as the basis for resimulation, a high degree of accuracy is fundamental. In this work, simplified scenarios on a testing site and a differential global positioning system (DGPS) with an inertial measurement unit as a reference system are used for precise ground truth acquisition. The sensor data was recorded in a scenario in which the ego vehicle is stationary and a target vehicle drives a path in the form of an eight in front of it. This scenario is well suited to analyze the influence of a spatial change of the observed object on the generation of radar detections.

\subsection{Synthetic Data Generation}
The generation of synthetic sensor data consists of two steps, namely the resimulation of real test drives and the actual generation of a virtual scene of the environment from the sensor point of view.

Contrary to the approach presented, it would be conceivable to simulate all possible scenarios not only those observed in real test drive. However, since a comparison to real data is crucial, a resimulation approach is used to overcome this issue. The determination of relevant scenarios that show the limits of sensor models is not scope of this present work. Both, the generated samples and the scenarios are utilized to reproduce the real test drives in simulation. The number of samples determines how often and with which parameter configuration a scenario is simulated.

The open-source simulator CARLA \cite{dosovitskiy_carla_2017} is utilized to implement the described procedure. Given the characteristics set out in Section \ref{sec2.1: Radar Sensor Modeling}, the ray tracing or rather ray casting method is used in this work to model the radar sensor. Especially with this approach, it is necessary to determine the sufficient level of detail of the individual sub-components in order to overcome the limitations in execution speed. The symbols used in the following equations, their units and descriptions are listed in Table \ref{tab: List of symbols}.

Radar is an electromagnetic system for the detection and location of reflecting objects and operates by radiating energy into space and detecting the echo signal reflected from an object. The radar range equation (also simply known as radar equation) relates the range of a radar sensor to the characteristics of the transmitter, target, environment, antenna, and the receiver. It is not only useful for estimating the maximum distance at which a particular radar can detect a target, but it can serve as a means for understanding the factors affecting radar performance \cite{skolnik_introduction_2001}.

\begin{equation} \label{eq: radar equation}
P_r = \frac{P_t G^2 \lambda^2 \sigma}{(4\pi)^3 R^4 L_{sys}}
\end{equation}

For this very reason, the radar equation is used in this work to model the signal power received $P_r$ and it can be calculated according to Equation \ref{eq: radar equation}. A detailed derivation of this equation can be found in \cite{skolnik_radar_2008}.

\begin{table}[h]
	\caption{List of symbols, their units and description}
	\label{tab: List of symbols}
	\begin{center}
		\begin{tabularx}{\linewidth}{ >{\centering\arraybackslash}p{1cm}
				>{\centering\arraybackslash}p{1cm}
				>{\raggedright\arraybackslash}X }
			\toprule
			Symbol & Unit & Description \\
			\midrule
			$B_n$		&		$\sfrac{1}{s}$	&		noise bandwidth  \\
			$G$		&		-				&		transmitting \& receiving antenna gain \\
			$k_B$			&		$\sfrac{J}{K}$	&		Boltzmann constant  \\
			$F_n$			&		-	&		noise figure  \\
			$L_{sys}$			&		-	&		overall system loss \\
			$P_t$		&		$W$				&		transmitting power \\
			$P_n$		&		$W$				&		noise power \\
			$P_r$		&		$W$				&		receiving power \\
			$R$			&		$m$				&		radial distance \\
			$SNR$		&		$-$				&		signal-to-noise ratio \\
			$T_0$	&		$K$				&		 standard temperature \\
			$\lambda$	&		$m$				&		wavelength of transmitted signal \\
			$\sigma$	&		$m^{2}$			&		radar cross-section \\
			\bottomrule
		\end{tabularx}
	\end{center}
\end{table}

\begin{figure}[thpb]
	\centering
	\setlength{\fboxrule}{0pt}
	\framebox{\parbox{\linewidth}{
			\includegraphics[scale=0.5] {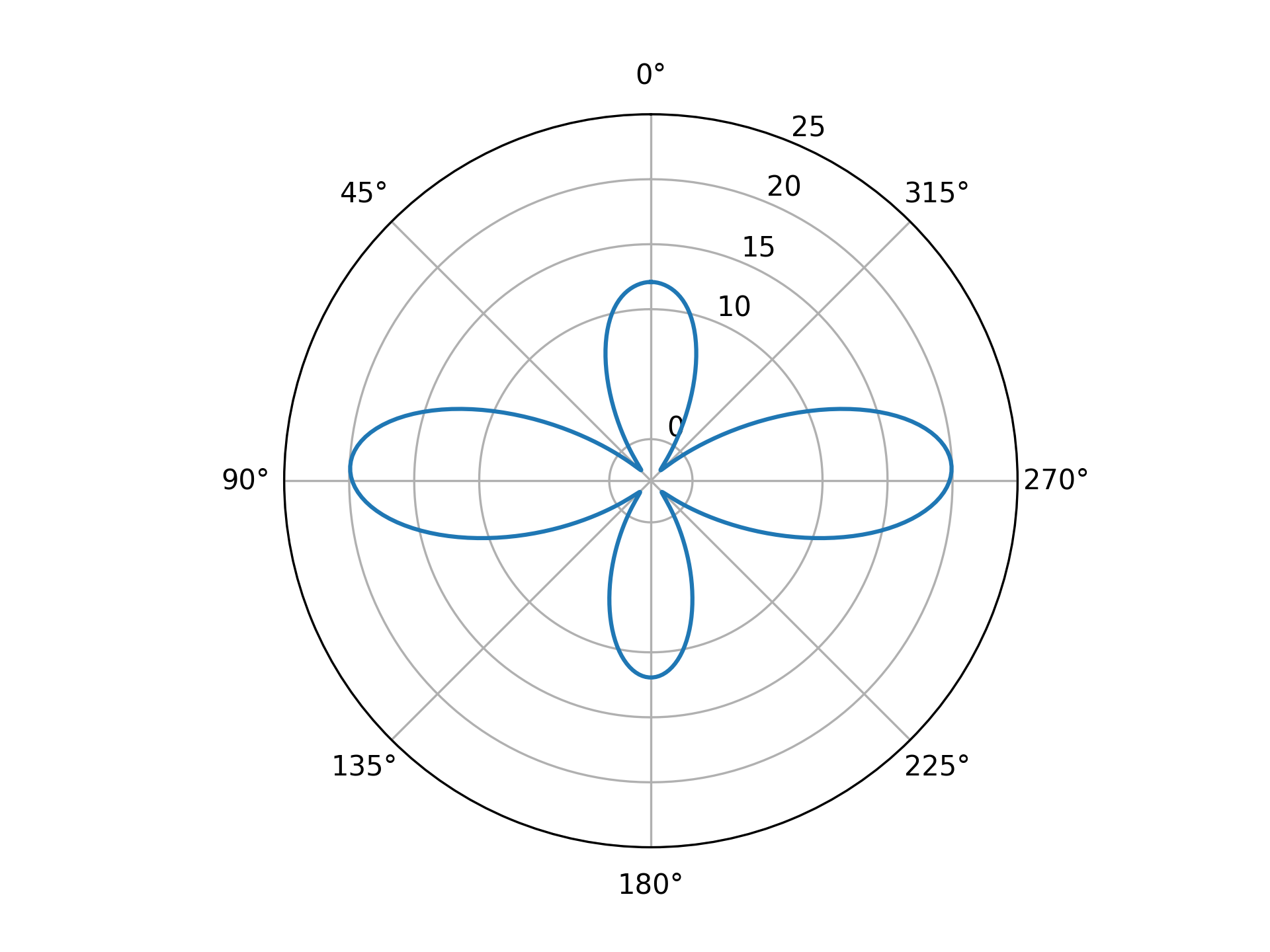}
	}}
	\caption{Radar cross-section}
	\label{fig: rcs}
\end{figure}

\begin{figure}[thpb]
	\centering
	\setlength{\fboxrule}{0pt}
	\framebox{\parbox{\linewidth}{
			\includegraphics[width=\linewidth] {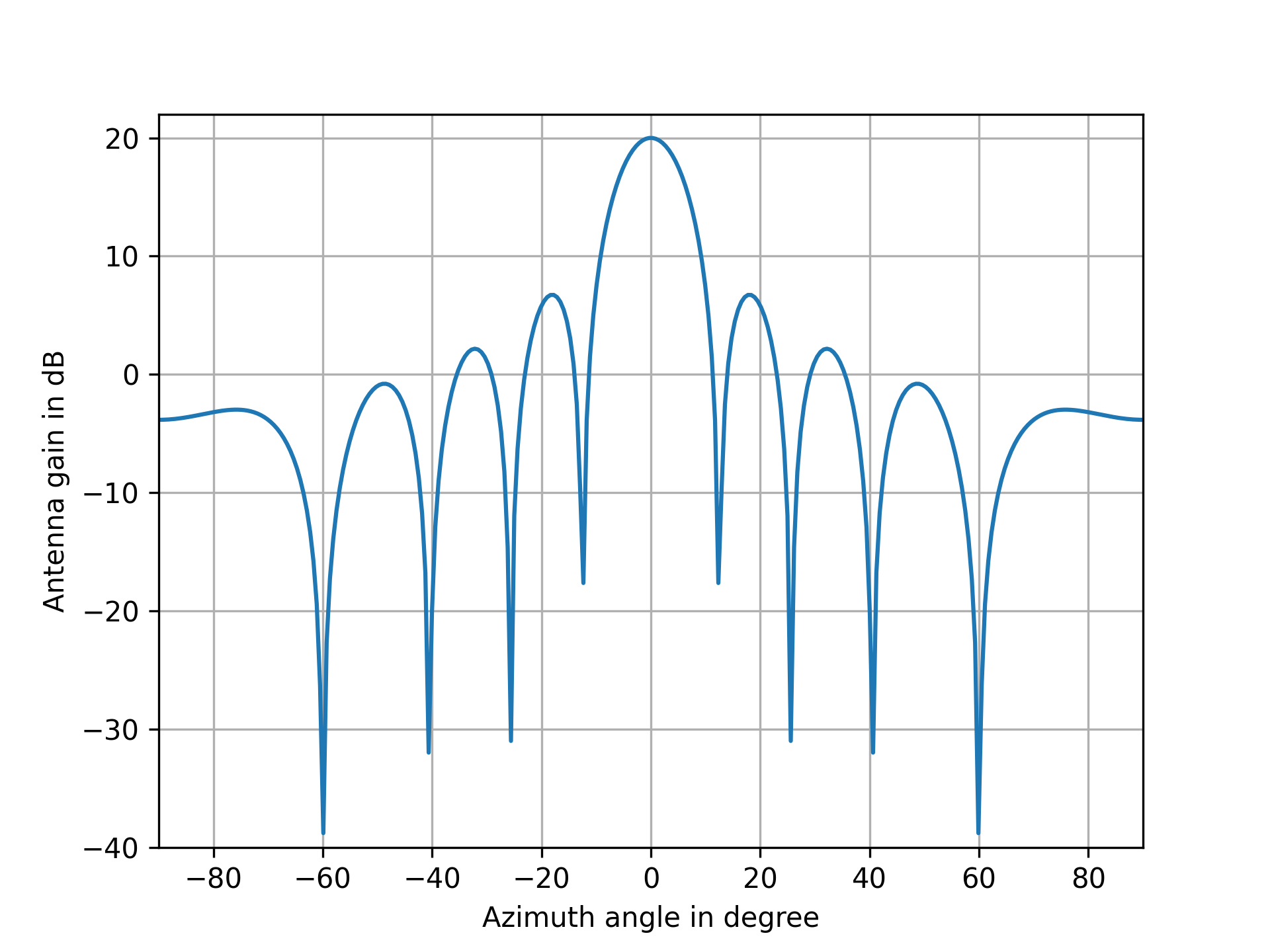}
	}}
	\caption{Antenna diagram}
	\label{fig: antenna gain}
\end{figure}

The radar cross-section $\sigma$ is assumed to be an ideal model depending on the aspect angle to the object (Fig. \ref{fig: rcs}). We focus in this work only on vehicles as objects and the corresponding radar cross-section (RCS) values are derived from the work of \cite{abadpour_extraction_2019} and \cite{matsunami_rcs_2012}. A generic antenna gain $G$ is used for both transmitting and receiving with a maximum gain of 20 dB and a side lobe suppression of -13 dB. Additionally, the antenna diagram is approximated with a simple sinc filter resulting in (Fig. \ref{fig: antenna gain}).

Furthermore, the ability of a radar sensor to detect an echo signal is limited by the ever-present noise that occupies the same part of the frequency spectrum as the radio signal. This is represented by the noise power $P_n$, which depends on the Boltzmann constant $k$,  the noise figure $F_n$, the noise bandwidth $B_n$ and the standard temperature $T_0$ (see Equation \ref{eq: noise power}). The noise power is modeled as additive white Gaussian noise (AWG).

\begin{equation} \label{eq: noise power}
P_n = k_B F_n B_n T_0
\end{equation}

In general, the performance of a radar's detection can be expressed by the ratio between the received signal power $P_r$ and the noise power $P_n$, resulting in the signal-to-noise ratio $SNR$ (Equation \ref{eq: snr short}).

\begin{equation} \label{eq: snr short}
SNR = \frac{P_r}{P_n}
\end{equation}

In combination with Equation \ref{eq: radar equation} and Equation \ref{eq: noise power} the signal-to-noise ratio of a radar sensor can be calculated as defined by the Equation \ref{eq: snr full}.

\begin{equation} \label{eq: snr full}
SNR = \frac{P_t G^2 \lambda^2 \sigma}	{k_B F_n B_n T_0 (4\pi)^3 R^4 L_{sys} }
\end{equation}

In order to generate detections from the reflected energy and the $SNR$ a threshold detection is applied. Hence, the detection of a radar signal is based on establishing a threshold at the output of the receiver. This threshold determines whether the receiver output is perceived as a detection present or as noise. Since noise is a random phenomenon, the detection of signals in the presence of noise is also a random phenomenon \cite{skolnik_introduction_2001}.
In the present work, the probabilistic behavior is incorporated with detection probabilities (DP). These probabilities can be determined in a simplified manner by a conversion of the signal-to-noise ratio via receiver operating curves (ROC) \cite{ponn_systematic_2019}. The ROCs are affected by the prevailing weather situation and can be dynamically adapted in real radar systems \cite{bernsteiner_radar_2015}, which leads to a shift in the SNR to detection probability conversion. For example, in rainy weather conditions the minimum SNR to generate a detection is usually increased to minimize false positives.

\subsection{System Under Test \& Evaluation}
After having collected real radar data as well as virtually generated radar data in the preceding steps, both serve as input for the SUT and thus form the basis for an evaluation. The system under test represents the subsequent stage after the sensor in the perception processing chain and is realized by a spatial clustering algorithm using radar detections. K-mean is a well known unsupervised learning algorithm and is used for clustering in this work. Furthermore, the algorithm is evaluated by comparing the euclidean distance of the predicted cluster centers. In this manner, the spatial distribution of the synthetically generated radar detections can be analyzed. The question as to which are the appropriate metrics to evaluate the simulation quality is still unresolved and is not the focus of this present paper.

\subsection{Parameters, Generation of Samples \& Sensitivity Analysis}
In the final step, a sensitivity analysis is conducted in order to determine the effect of individual factors in driving the output and its uncertainty. Sensitivity analysis is the study of how uncertainty in the output of a model can be apportioned to different sources of uncertainty in the model input \cite{saltelli_global_2008}. Whereas, by contrast, uncertainty analysis focuses on quantifying the uncertainty in model output. Examples of applications of sensitivity analysis are model simplification in the context of complex and computer demanding models, quality assurance or robust assessment \cite{saltelli_sensitivity_2011}.

Sensitivity analysis methods can be inter alia differentiated into qualitative and quantitative methods, while in general qualitative methods are more efficient, but less accurate. Furthermore, Fourier amplitude sensitivity testing (FAST), which was presented by \cite{cukier_study_1973}, is an effective variance-based quantitative sensitivity analysis method and is used in this present work. Variance-based methods provide quantitative measures of how much each varying parameters contributes to the overall variance of the model response as well as quantifying the interaction effect of the parameters. They can be applied to complex non-linear and non-monotonic model \cite{gan_comprehensive_2014}. 
% input 
The parameters listed in Table \ref{tab: Sensitivity analysis parameters} are the input parameters for the FAST method in order to analyze the impact of the different sensor effects on the result of the SUT. The bounds of the parameters are designed according to typical automotive radar values as reported by \cite{gamba_radar_2020} and \cite{skolnik_radar_2008}. The parameters are varied within these specified bounds to analyze the corresponding output variance.
% output
The output of the FAST method are both first-order sensitivity index $S_i$ and total-order sensitivity index $S_{Ti}$. The first-order sensitivity index quantifies the main effect of a parameter, whereas the total-order sensitivity index describes the overall effect of a parameter, i.e. the total effect. The difference $S_{Ti} - S_i$ is a measure of the strength of the interactions \cite{saltelli_sensitivity_2011}.

\begin{table}[h]
	\caption{Sensitivity analysis parameters and their bounds}
	\label{tab: Sensitivity analysis parameters}
	\begin{center}
				\begin{tabular}{l l l c l} % use tabular when width 
%		\begin{tabular}{\linewidth}{ >{\centering\arraybackslash}p{1cm}
%				>{\centering\arraybackslash}p{1cm}
%				>{\right\arraybackslash}X }
%				>{\right\arraybackslash}X }
			\toprule
			Symbol & Description & Unit & Min/Max \\
			\midrule
				$AWGNoise$	&	AWG noise standard deviation	&	$dB$	  &		$0 / 8$\\
				$DP_{offset}$	&	detection probability offset	&	$-$	&		$-5./ 5$\\
				$G_{max}$	& maximal antenna gain	&		$dB$				&	$10 / 25$ \\
				$F_n$		& noise figure	&	$dB$	&		$10 / 20$  \\
				$L_{sys}$	&	overall system loss	&	$dB$	&		$0 / 20$\\
				$RCS_{mean}$ &	mean radar cross-section  &		$dBsm$	& $-10 / 10$ \\
			\bottomrule
		\end{tabular}
	\end{center}
\end{table}

%%%%%%%%%%%%%%%%%%%%%%%%%%%%%%%%%%%%%%
\section{EXPERIMENTS AND RESULTS} \label{sec4: Results}
This section evaluates the results of the method presented in Section \ref{sec3: Method} and discusses the effectiveness of the proposed method.

\subsection{Clustering Evaluation}
\begin{figure}[thpb]
	\centering
	%	\framebox{\parbox{3in}{We suggest that you use a text box to insert a graphic (which is ideally a 300 dpi TIFF or EPS file, with all fonts embedded) because, in an document, this method is somewhat more stable than directly inserting a picture.
	%	}}
	% so that width is same as text: width=\textwidth
%	\includegraphics[width=\textwidth] {./figures/clustering_results.png}
	\includegraphics[scale=0.61] {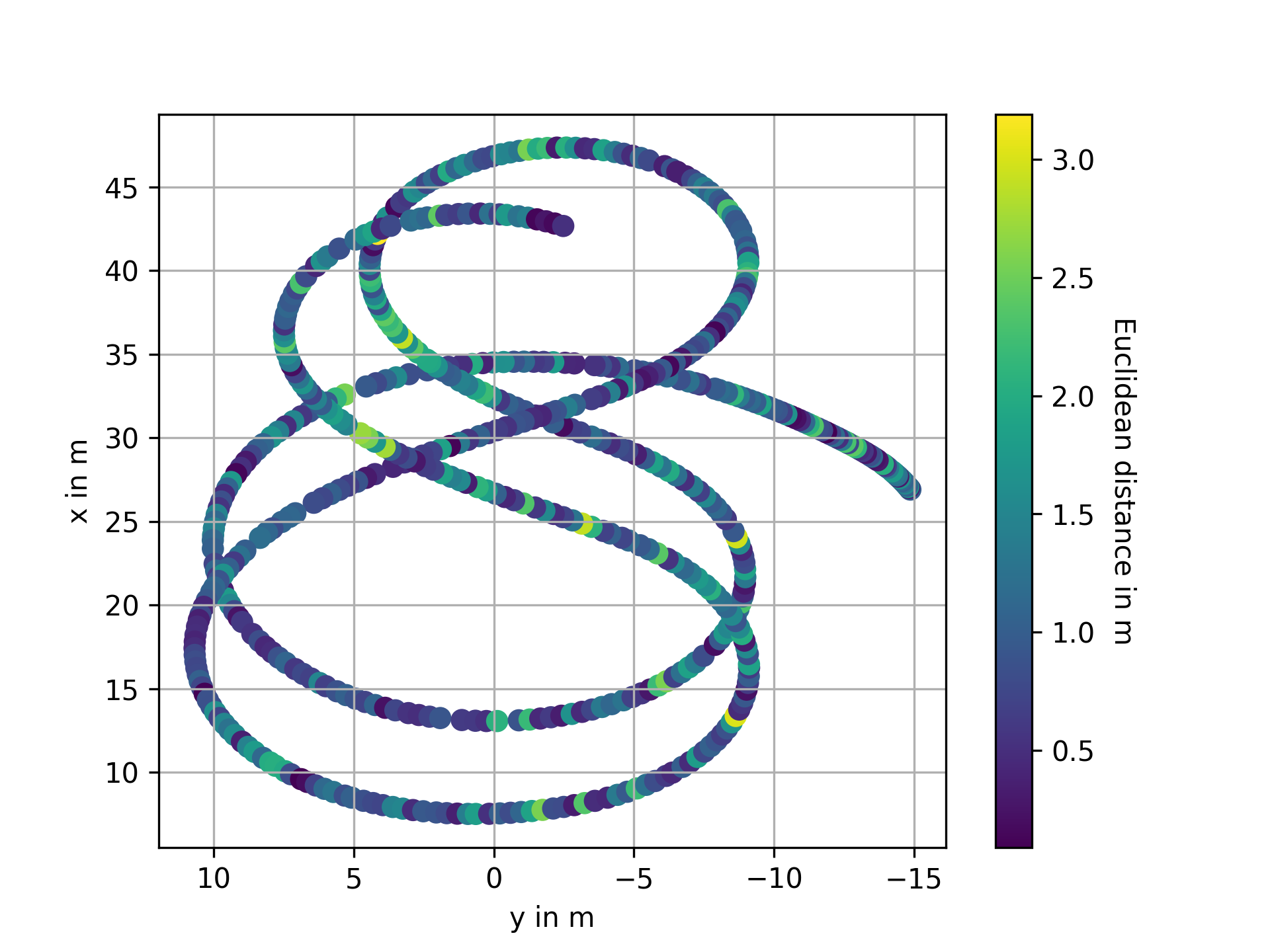}
	\caption{Clustering evaluation results}
	\label{fig: clustering results}
\end{figure}

In the experiment a target object drives a path in the form of an eight in front of the radar sensor, which itself is stationary. The objective in this scenario is to analyze the influence of the orientation of the object on the generation of radar detections from different distances to the sensor. It can be assumed that the detections change in density and distribution over the range. In this first evaluation it will be investigated whether and to what extent the radar sensor model can approximate this behavior.
For the purpose of investigating this, the clustering algorithm is applied with both real and synthetically generated radar sensor data. As a simple metric the euclidean distance is computed between both predicted cluster centroids, resulting in Fig. \ref{fig: clustering results}. 

Obviously, deviations between simulation and measurement can be observed. Particularly noticeable is the relatively larger diversion when the object turns. This is probably due to an inaccurate antenna model and/or RCS model. Knowing that the object starts from the negative y-axis, it can be seen that larger differences occur above all when the object is observed from the rear view. This indicates that a symmetric RCS model is not sufficient, at least for the longitudinal vehicle axis.

\subsection{Sensitivity Analysis}
\begin{figure}[thpb]
	\centering
	%	\framebox{\parbox{3in}{We suggest that you use a text box to insert a graphic (which is ideally a 300 dpi TIFF or EPS file, with all fonts embedded) because, in an document, this method is somewhat more stable than directly inserting a picture.
	%	}}
	% so that width is same as text: width=\textwidth
	\includegraphics[scale=0.57] {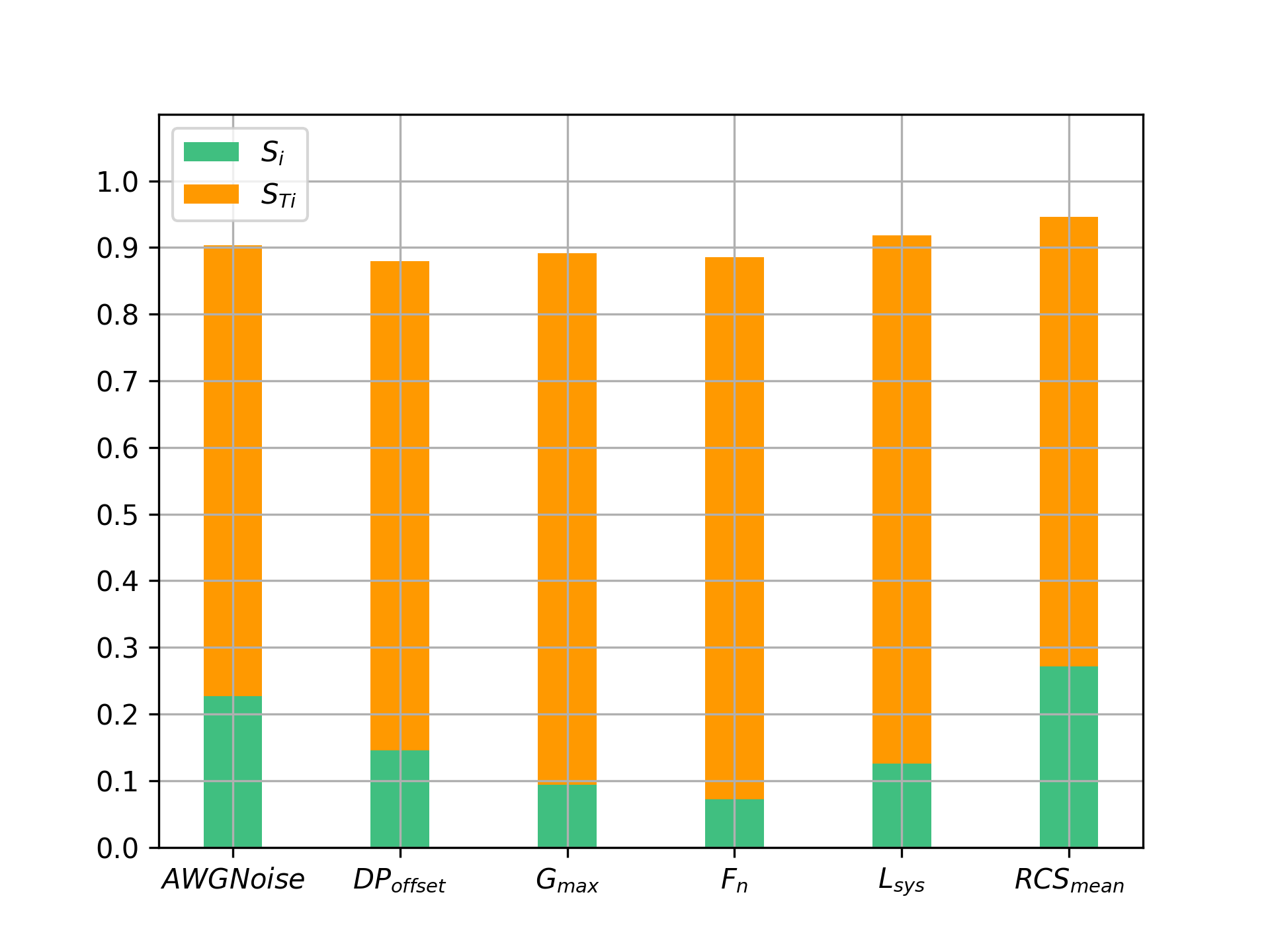}
	\caption{Sensitivity analysis results using minimum evaluation values}
	\label{fig: sensitivity result min}
\end{figure}

\begin{figure}[thpb]
	\centering
	%	\framebox{\parbox{3in}{We suggest that you use a text box to insert a graphic (which is ideally a 300 dpi TIFF or EPS file, with all fonts embedded) because, in an document, this method is somewhat more stable than directly inserting a picture.
	%	}}
	% so that width is same as text: width=\textwidth
	\includegraphics[scale=0.57] {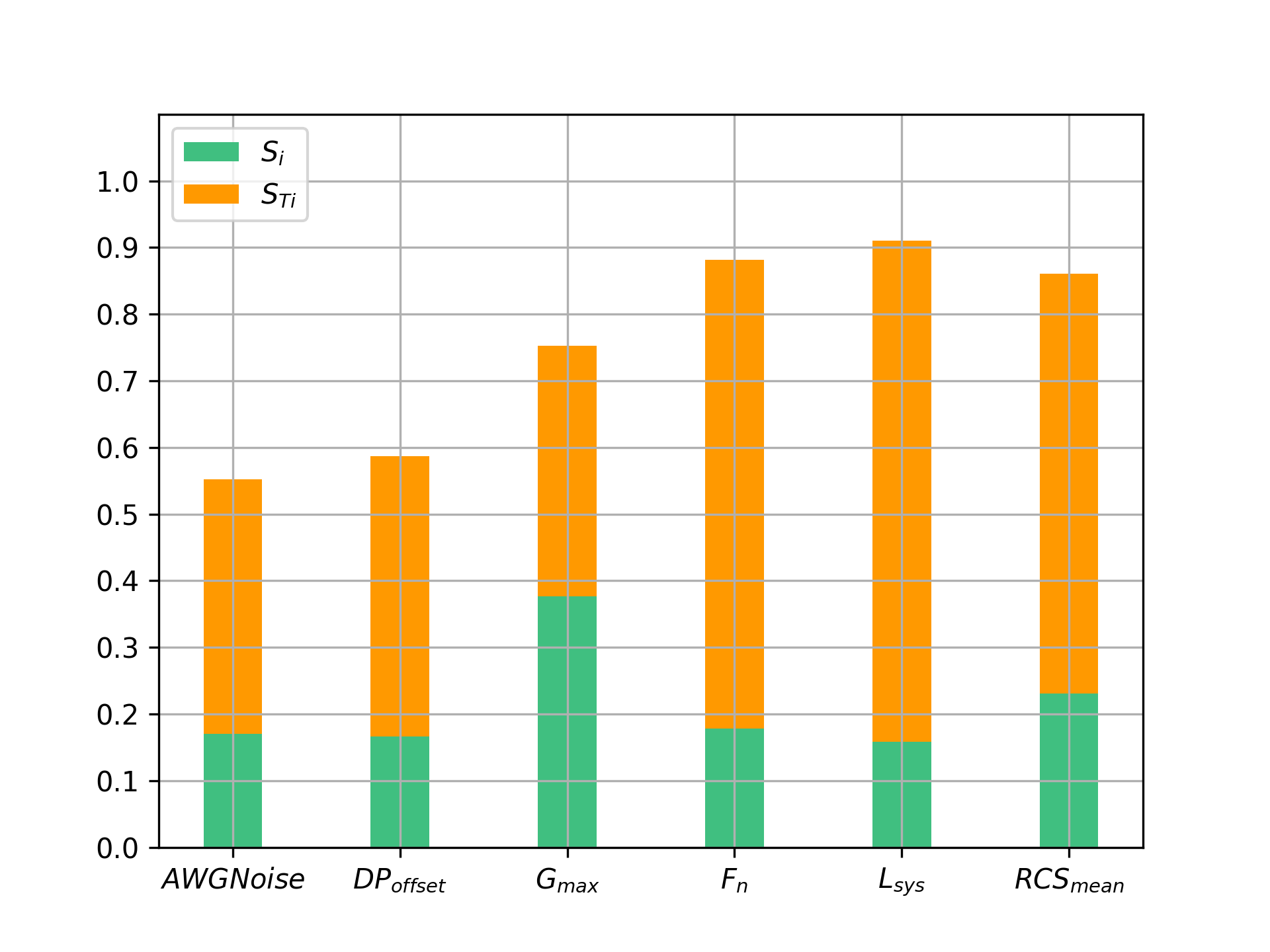}
	\caption{Sensitivity analysis results using mean evaluation values}
	\label{fig: sensitivity result mean}
\end{figure}

\begin{figure}[thpb]
	\centering
	%	\framebox{\parbox{3in}{We suggest that you use a text box to insert a graphic (which is ideally a 300 dpi TIFF or EPS file, with all fonts embedded) because, in an document, this method is somewhat more stable than directly inserting a picture.
	%	}}
	% so that width is same as text: width=\textwidth
	\includegraphics[scale=0.57] {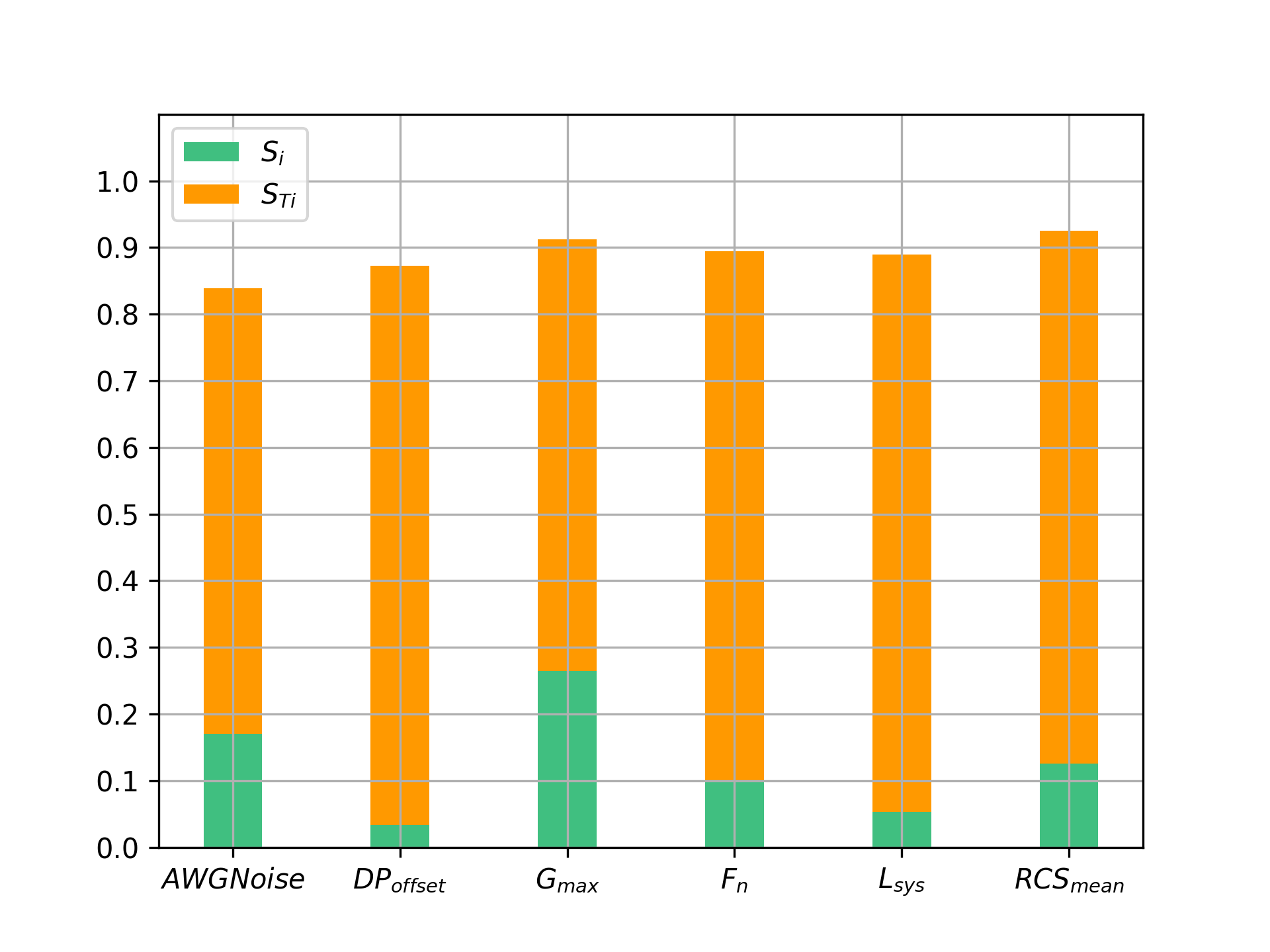}
	\caption{Sensitivity analysis results using maximum evaluation values}
	\label{fig: sensitivity result max}
\end{figure}

This section covers the results of the sensitivity analysis. The radar sensor model parameter are parameterized according to Table \ref{tab: Sensitivity analysis parameters} in order to investigate the impact of each parameter regarding the clustering algorithm as the system under test. The parameters are varied within the specified bounds with the purpose of creating the samples, leading to a total of 390 samples used in this work. Each sample reflects a certain parameter configuration, which is then simulated to generate the synthetic radar sensor data. These sensor data are evaluated analogously to the evaluation approach described previously to generate the input for the FAST method. For the reason that the given sensitivity method requires only one scalar value per simulation run as input, the sensitivity analysis is conducted with three different approaches to calculate the evaluation value: minimum, mean and maximum of the euclidian distance over all frames of a simulation run. Consequently, the results of these three sensitivity analyses are illustrated in Fig. \ref{fig: sensitivity result min}, \ref{fig: sensitivity result mean} and \ref{fig: sensitivity result max}.

It is evident that no parameter alone is mainly responsible for the result, this can be concluded from the first-order sensitivity indices $Si$ of each evaluation. However, no parameter can be neglected due to the consistently high interaction coefficient, which is calculated by $S_{Ti} - S_i$ \cite{gan_comprehensive_2014}. The main effect of both the noise figure and the standard deviation of the AWG noise remain in the same range of $10-20\%$. The total loss is relatively high for both the mean and the max evaluation, which leads to the assumption that maximum bound of this parameter too large. From the evaluation results from the previous section, it can be assumed that the minimal deviation occur especially at small distances. And vice versa, that the large discrepancies between simulation and measurement are rather at long distances. Given this assumption, the relatively high $S_i$ of the detection probability for the minimum case can be traced back to a higher number of radar detections in the sensor near field due to the applied ray casting method. A relatively constant high influence of the antenna is to be expected, since this depends mainly on the azimuth angle. In contrast, the main effect of the RCS model has the least impact in the maximum case, i.e. at large distances. This  indicates that a pure aspect angle dependent rcs model is not sufficient.

\section{CONCLUSION} \label{sec5: Conclusion}
In this paper, a method for evaluating a radar sensor model was presented to determine the impact of sensor effects regarding a given system under test. A clustering algorithm was used as a system under test, which processes both synthetic sensor data and real sensor data. In order to investigate the effectiveness of the proposed method the FAST algorithm was utilized to conduct a sensitivity analysis taking the result of the clustering evaluation as input. 
It was shown that a sensitivity analysis enables a more detailed evaluation of the synthetic generated sensor data. The results from different situations can be traced back to the contribution of the individual sub-modules or sensor effects of the radar sensor model. This leads to an efficient analysis of the simulation result. The developed
method can complement the research towards virtual validation of autonomous driving functions.
% further research
There are several conceivable extensions for the approach to further enhance the evaluation. In addition to radar sensor effects, simulation-related parameters can be examined, such as the number of reflections during ray tracing or the number of emitted rays. The required computing time must also be considered, which increases exponentially with a higher number of reflections.

% set this to zero if there are a lot of references
\addtolength{\textheight}{0cm}   % This command serves to balance the column lengths


% References are important to the reader; therefore, each citation must be complete and correct. If at all possible, references should be commonly available publications.

\bibliography{MyLibrary}

\begin{thebibliography}{10}
\providecommand{\url}[1]{#1}
\csname url@rmstyle\endcsname
\providecommand{\newblock}{\relax}
\providecommand{\bibinfo}[2]{#2}
\providecommand\BIBentrySTDinterwordspacing{\spaceskip=0pt\relax}
\providecommand\BIBentryALTinterwordstretchfactor{4}
\providecommand\BIBentryALTinterwordspacing{\spaceskip=\fontdimen2\font plus
\BIBentryALTinterwordstretchfactor\fontdimen3\font minus
  \fontdimen4\font\relax}
\providecommand\BIBforeignlanguage[2]{{%
\expandafter\ifx\csname l@#1\endcsname\relax
\typeout{** WARNING: IEEEtran.bst: No hyphenation pattern has been}%
\typeout{** loaded for the language `#1'. Using the pattern for}%
\typeout{** the default language instead.}%
\else
\language=\csname l@#1\endcsname
\fi
#2}}

\bibitem{friedrich_effect_2016}
B.~Friedrich, ``The {Effect} of {Autonomous} {Vehicles} on {Traffic},'' in
  \emph{Autonomous {Driving}: {Technical}, {Legal} and {Social} {Aspects}},
  M.~Maurer, J.~C. Gerdes, B.~Lenz, and H.~Winner, Eds.\hskip 1em plus 0.5em
  minus 0.4em\relax Berlin, Heidelberg: Springer Berlin Heidelberg, 2016, pp.
  317--334.

\bibitem{koopman_challenges_2016}
\BIBentryALTinterwordspacing
P.~Koopman and M.~Wagner, ``\BIBforeignlanguage{en}{Challenges in {Autonomous}
  {Vehicle} {Testing} and {Validation}},'' \emph{\BIBforeignlanguage{en}{SAE
  International Journal of Transportation Safety}}, vol.~4, no.~1, pp. 15--24,
  Apr. 2016. [Online]. Available:
  \url{https://www.sae.org/content/2016-01-0128/}
\BIBentrySTDinterwordspacing

\bibitem{wachenfeld_release_2016}
W.~Wachenfeld and H.~Winner, ``The {Release} of {Autonomous} {Vehicles},'' in
  \emph{Autonomous {Driving}: {Technical}, {Legal} and {Social} {Aspects}},
  M.~Maurer, J.~C. Gerdes, B.~Lenz, and H.~Winner, Eds.\hskip 1em plus 0.5em
  minus 0.4em\relax Berlin, Heidelberg: Springer Berlin Heidelberg, 2016, pp.
  425--449.

\bibitem{jesenski_simulation-based_2019}
\BIBentryALTinterwordspacing
S.~Jesenski, J.~E. Stellet, W.~Branz, and J.~M. Zollner,
  ``\BIBforeignlanguage{en}{Simulation-{Based} {Methods} for {Validation} of
  {Automated} {Driving}: {A} {Model}-{Based} {Analysis} and an {Overview} about
  {Methods} for {Implementation}},'' in \emph{\BIBforeignlanguage{en}{2019
  {IEEE} {Intelligent} {Transportation} {Systems} {Conference}
  ({ITSC})}}.\hskip 1em plus 0.5em minus 0.4em\relax Auckland, New Zealand:
  IEEE, Oct. 2019, pp. 1914--1921. [Online]. Available:
  \url{https://ieeexplore.ieee.org/document/8917072/}
\BIBentrySTDinterwordspacing

\bibitem{schaermann_validation_2017}
\BIBentryALTinterwordspacing
A.~Schaermann, A.~Rauch, N.~Hirsenkorn, T.~Hanke, R.~Rasshofer, and E.~Biebl,
  ``\BIBforeignlanguage{en}{Validation of vehicle environment sensor models},''
  in \emph{\BIBforeignlanguage{en}{2017 {IEEE} {Intelligent} {Vehicles}
  {Symposium} ({IV})}}.\hskip 1em plus 0.5em minus 0.4em\relax Los Angeles, CA,
  USA: IEEE, June 2017, pp. 405--411. [Online]. Available:
  \url{http://ieeexplore.ieee.org/document/7995752/}
\BIBentrySTDinterwordspacing

\bibitem{oberkampf_simulation_2019}
W.~L. Oberkampf, ``Simulation {Accuracy}, {Uncertainty}, and {Predictive}
  {Capability}: {A} {Physical} {Sciences} {Perspective},'' in \emph{Computer
  {Simulation} {Validation}: {Fundamental} {Concepts}, {Methodological}
  {Frameworks}, and {Philosophical} {Perspectives}}, C.~Beisbart and N.~J.
  Saam, Eds.\hskip 1em plus 0.5em minus 0.4em\relax Cham: Springer
  International Publishing, 2019, pp. 69--97.

\bibitem{wheeler_deep_2017}
\BIBentryALTinterwordspacing
T.~A. Wheeler, M.~Holder, H.~Winner, and M.~Kochenderfer,
  ``\BIBforeignlanguage{en}{Deep {Stochastic} {Radar} {Models}},''
  \emph{\BIBforeignlanguage{en}{arXiv:1701.09180 [cs]}}, June 2017, arXiv:
  1701.09180. [Online]. Available: \url{http://arxiv.org/abs/1701.09180}
\BIBentrySTDinterwordspacing

\bibitem{chipengo_antenna_2018}
\BIBentryALTinterwordspacing
U.~Chipengo, P.~M. Krenz, and S.~Carpenter, ``\BIBforeignlanguage{en}{From
  {Antenna} {Design} to {High} {Fidelity}, {Full} {Physics} {Automotive}
  {Radar} {Sensor} {Corner} {Case} {Simulation}},''
  \emph{\BIBforeignlanguage{en}{Modelling and Simulation in Engineering}}, vol.
  2018, pp. 1--19, Dec. 2018. [Online]. Available:
  \url{https://www.hindawi.com/journals/mse/2018/4239725/}
\BIBentrySTDinterwordspacing

\bibitem{skolnik_radar_2008}
M.~I. Skolnik, \emph{\BIBforeignlanguage{en}{Radar {Handbook}}}, 3rd~ed.\hskip
  1em plus 0.5em minus 0.4em\relax New York, USA: McGraw-Hill, 2008.

\bibitem{yun_ray_2015}
\BIBentryALTinterwordspacing
Z.~Yun and M.~F. Iskander, ``\BIBforeignlanguage{en}{Ray {Tracing} for {Radio}
  {Propagation} {Modeling}: {Principles} and {Applications}},''
  \emph{\BIBforeignlanguage{en}{IEEE Access}}, vol.~3, pp. 1089--1100, 2015.
  [Online]. Available: \url{http://ieeexplore.ieee.org/document/7152831/}
\BIBentrySTDinterwordspacing

\bibitem{yee_numerical_1966}
\BIBentryALTinterwordspacing
K.~Yee, ``\BIBforeignlanguage{en}{Numerical solution of initial boundary value
  problems involving maxwell's equations in isotropic media},''
  \emph{\BIBforeignlanguage{en}{IEEE Transactions on Antennas and
  Propagation}}, vol.~14, no.~3, pp. 302--307, May 1966. [Online]. Available:
  \url{http://ieeexplore.ieee.org/document/1138693/}
\BIBentrySTDinterwordspacing

\bibitem{clemens_discrete_2001}
\BIBentryALTinterwordspacing
M.~Clemens and T.~Weiland, ``\BIBforeignlanguage{en}{Discrete
  {Electromagnetism} with the {Finite} {Integration} {Technique}},''
  \emph{\BIBforeignlanguage{en}{Progress In Electromagnetics Research}},
  vol.~32, pp. 65--87, 2001. [Online]. Available:
  \url{http://www.jpier.org/PIER/pier.php?paper=00080103}
\BIBentrySTDinterwordspacing

\bibitem{jin_finite_2014}
J.-M. Jin, \emph{The finite element method in electromagnetics}, third
  edition~ed.\hskip 1em plus 0.5em minus 0.4em\relax Hoboken. New Jersey: John
  Wiley \& Sons Inc, 2014.

\bibitem{harrington_field_1968}
\BIBentryALTinterwordspacing
R.~Harrington, \emph{Field computation by moment methods}, ser. Macmillan
  series in electrical science.\hskip 1em plus 0.5em minus 0.4em\relax
  Macmillan, 1968. [Online]. Available:
  \url{https://books.google.de/books?id=tgNRAAAAMAAJ}
\BIBentrySTDinterwordspacing

\bibitem{machida_rapid_2019}
\BIBentryALTinterwordspacing
T.~Machida and T.~Owaki, ``\BIBforeignlanguage{en}{Rapid and {Precise}
  {Millimeter}-wave {Radar} {Simulation} for {ADAS} {Virtual} {Assessment}},''
  in \emph{\BIBforeignlanguage{en}{2019 {IEEE} {Intelligent} {Transportation}
  {Systems} {Conference} ({ITSC})}}.\hskip 1em plus 0.5em minus 0.4em\relax
  Auckland, New Zealand: IEEE, Oct. 2019, pp. 431--436. [Online]. Available:
  \url{https://ieeexplore.ieee.org/document/8917498/}
\BIBentrySTDinterwordspacing

\bibitem{owaki_hybrid_2019}
\BIBentryALTinterwordspacing
T.~Owaki and T.~Machida, ``\BIBforeignlanguage{en}{Hybrid {Physics}-{Based} and
  {Data}-{Driven} {Approach} to {Estimate} the {Radar} {Cross}-{Section} of
  {Vehicles}},'' in \emph{\BIBforeignlanguage{en}{2019 {IEEE} {Intelligent}
  {Transportation} {Systems} {Conference} ({ITSC})}}.\hskip 1em plus 0.5em
  minus 0.4em\relax Auckland, New Zealand: IEEE, Oct. 2019, pp. 673--678.
  [Online]. Available: \url{https://ieeexplore.ieee.org/document/8917492/}
\BIBentrySTDinterwordspacing

\bibitem{keller_geometrical_1962}
\BIBentryALTinterwordspacing
J.~B. Keller, ``\BIBforeignlanguage{EN}{Geometrical {Theory} of
  {Diffraction}},'' \emph{\BIBforeignlanguage{EN}{Journal of the Optical
  Society of America}}, vol.~52, no.~2, pp. 116--130, Feb. 1962, publisher:
  Optical Society of America. [Online]. Available:
  \url{https://www.osapublishing.org/josa/abstract.cfm?uri=josa-52-2-116}
\BIBentrySTDinterwordspacing

\bibitem{holder_measurements_2018}
\BIBentryALTinterwordspacing
M.~Holder, P.~Rosenberger, H.~Winner, T.~Dhondt, V.~P. Makkapati, M.~Maier,
  H.~Schreiber, Z.~Magosi, Z.~Slavik, O.~Bringmann, and W.~Rosenstiel,
  ``\BIBforeignlanguage{en}{Measurements revealing {Challenges} in {Radar}
  {Sensor} {Modeling} for {Virtual} {Validation} of {Autonomous} {Driving}},''
  in \emph{\BIBforeignlanguage{en}{International {Conference} on {Intelligent}
  {Transportation} {Systems} ({ITSC})}}.\hskip 1em plus 0.5em minus 0.4em\relax
  Maui, Hawaii, USA: IEEE, Nov. 2018, pp. 2616--2622. [Online]. Available:
  \url{https://ieeexplore.ieee.org/document/8569423/}
\BIBentrySTDinterwordspacing

\bibitem{cao_modeling_2017}
P.~Cao, ``\BIBforeignlanguage{en}{Modeling {Active} {Perception} {Sensors} for
  {Real}-{Time} {Virtual} {Validation} of {Automated} {Driving} {Systems}},''
  Ph.{D}. dissertation, Technical University of Darmstadt, Darmstadt, Germany,
  2017.

\bibitem{martowicz_uncertainty_2019}
\BIBentryALTinterwordspacing
A.~Martowicz, A.~Gallina, and G.~Karpiel, ``\BIBforeignlanguage{en}{Uncertainty
  propagation for vehicle detections in experimentally validated radar model
  for automotive application},'' in \emph{\BIBforeignlanguage{en}{2019 24th
  {International} {Conference} on {Methods} and {Models} in {Automation} and
  {Robotics} ({MMAR})}}.\hskip 1em plus 0.5em minus 0.4em\relax Miedzyzdroje,
  Poland: IEEE, Aug. 2019, pp. 606--611. [Online]. Available:
  \url{https://ieeexplore.ieee.org/document/8864641/}
\BIBentrySTDinterwordspacing

\bibitem{roth_analysis_2011}
E.~Roth, T.~J. Dirndorfer, A.~Knoll, K.~v.~Neumann-Cosel, T.~Ganslmeier,
  A.~Kern, and M.-O. Fischer, ``Analysis and {Validation} of {Perception}
  {Sensor} {Models} in an {Integrated} {Vehicle} and {Environment}
  {Simulation},'' in \emph{22nd {Enhanced} {Safety} of {Vehicle} {Conference}},
  Washington DC, USA, 2011.

\bibitem{suhre_simulating_2018}
\BIBentryALTinterwordspacing
A.~Suhre and W.~Malik, ``\BIBforeignlanguage{en}{Simulating object lists using
  neural networks in automotive radar},'' in
  \emph{\BIBforeignlanguage{en}{International {Conference} on {Thermal},
  {Mechanical} and {Multi}-{Physics} {Simulation} and {Experiments} in
  {Microelectronics} and {Microsystems} ({EuroSimE})}}.\hskip 1em plus 0.5em
  minus 0.4em\relax Toulouse: IEEE, Apr. 2018, pp. 1--5. [Online]. Available:
  \url{https://ieeexplore.ieee.org/document/8369885/}
\BIBentrySTDinterwordspacing

\bibitem{rosenberger_towards_2019}
P.~Rosenberger, J.~T. Wendler, M.~Holder, C.~Linnhoff, M.~Berghoefer,
  H.~Winner, and M.~Maurer, ``\BIBforeignlanguage{en}{Towards a {Generally}
  {Accepted} {Validation} {Methodology} for {Sensor} {Models} - {Challenges},
  {Metrics}, and {First} {Results}},'' in \emph{\BIBforeignlanguage{en}{Graz
  {Symposium} {Virtual} {Vehicle} ({GSVF}) 2019}}, Graz, Austria, 2019, p.~13.

\bibitem{dosovitskiy_carla_2017}
A.~Dosovitskiy, G.~Ros, F.~Codevilla, A.~Lopez, and V.~Koltun,
  ``\BIBforeignlanguage{en}{{CARLA}: {An} {Open} {Urban} {Driving}
  {Simulator}},'' in \emph{\BIBforeignlanguage{en}{Proceedings of the 1st
  {Annual} {Conference} on {Robot} {Learning}}}, Mountain View, United States,
  2017, pp. 1--16.

\bibitem{skolnik_introduction_2001}
M.~I. Skolnik, \emph{\BIBforeignlanguage{en}{Introduction to {Radar}
  {Systems}}}, 3rd~ed.\hskip 1em plus 0.5em minus 0.4em\relax Boston, MA, USA:
  McGraw-Hill, 2001.

\bibitem{abadpour_extraction_2019}
\BIBentryALTinterwordspacing
S.~Abadpour, A.~Diewald, M.~Pauli, and T.~Zwick,
  ``\BIBforeignlanguage{en}{Extraction of {Scattering} {Centers} {Using} a 77
  {GHz} {FMCW} {Radar}},'' in \emph{\BIBforeignlanguage{en}{2019 12th {German}
  {Microwave} {Conference} ({GeMiC})}}.\hskip 1em plus 0.5em minus 0.4em\relax
  Stuttgart, Germany: IEEE, Mar. 2019, pp. 79--82. [Online]. Available:
  \url{https://ieeexplore.ieee.org/document/8698144/}
\BIBentrySTDinterwordspacing

\bibitem{matsunami_rcs_2012}
\BIBentryALTinterwordspacing
I.~Matsunami, R.~Nakamura, and A.~Kajiwara, ``\BIBforeignlanguage{en}{{RCS}
  measurements for vehicles and pedestrian at 26 and {79GHz}},'' in
  \emph{\BIBforeignlanguage{en}{2012 6th {International} {Conference} on
  {Signal} {Processing} and {Communication} {Systems}}}.\hskip 1em plus 0.5em
  minus 0.4em\relax Gold Coast, Australia: IEEE, Dec. 2012, pp. 1--4. [Online].
  Available: \url{http://ieeexplore.ieee.org/document/6508004/}
\BIBentrySTDinterwordspacing

\bibitem{ponn_systematic_2019}
\BIBentryALTinterwordspacing
T.~Ponn, F.~Mueller, and F.~Diermeyer, ``\BIBforeignlanguage{en}{Systematic
  {Analysis} of the {Sensor} {Coverage} of {Automated} {Vehicles} {Using}
  {Phenomenological} {Sensor} {Models}},'' in
  \emph{\BIBforeignlanguage{en}{2019 {IEEE} {Intelligent} {Vehicles}
  {Symposium} ({IV})}}.\hskip 1em plus 0.5em minus 0.4em\relax Paris, France:
  IEEE, June 2019, pp. 1000--1006. [Online]. Available:
  \url{https://ieeexplore.ieee.org/document/8813794/}
\BIBentrySTDinterwordspacing

\bibitem{bernsteiner_radar_2015}
S.~Bernsteiner, Z.~Magosi, D.~Lindvai-Soos, and A.~Eichberger,
  ``\BIBforeignlanguage{en}{Radar {Sensor} {Model} for the {Virtual}
  {Development} {Process}},'' \emph{\BIBforeignlanguage{en}{ATZelectronics
  worldwide}}, vol.~10, no.~2, pp. 46--52, Apr. 2015.

\bibitem{saltelli_global_2008}
A.~Saltelli, M.~Ratto, T.~Andres, and F.~Campolongo,
  \emph{\BIBforeignlanguage{en}{Global {Sensitivity} {Analysis}: {The}
  {Primer}}}.\hskip 1em plus 0.5em minus 0.4em\relax Chichester: Wiley, 2008.

\bibitem{saltelli_sensitivity_2011}
A.~Saltelli and P.~Annoni, ``\BIBforeignlanguage{en}{Sensitivity {Analysis}},''
  in \emph{\BIBforeignlanguage{en}{International {Encyclopedia} of
  {Statistical} {Science}}}.\hskip 1em plus 0.5em minus 0.4em\relax Springer
  Berlin Heidelberg, 2011.

\bibitem{cukier_study_1973}
R.~I. Cukier, C.~M. Fortuin, K.~E. Schuler, A.~G. Petschek, and J.~H. Schaibly,
  ``Study of the sensitivity of coupled reaction systems to uncertainties in
  rate coefficients,'' \emph{The Journal of Chemical Physics}, vol.~59, no.~8,
  pp. 3873--3878, 1973.

\bibitem{gan_comprehensive_2014}
\BIBentryALTinterwordspacing
Y.~Gan, Q.~Duan, W.~Gong, C.~Tong, Y.~Sun, W.~Chu, A.~Ye, C.~Miao, and Z.~Di,
  ``\BIBforeignlanguage{en}{A comprehensive evaluation of various sensitivity
  analysis methods: {A} case study with a hydrological model},''
  \emph{\BIBforeignlanguage{en}{Environmental Modelling \& Software}}, vol.~51,
  pp. 269--285, Jan. 2014. [Online]. Available:
  \url{https://linkinghub.elsevier.com/retrieve/pii/S1364815213002338}
\BIBentrySTDinterwordspacing

\bibitem{gamba_radar_2020}
\BIBentryALTinterwordspacing
J.~Gamba, \emph{\BIBforeignlanguage{en}{Radar {Signal} {Processing} for
  {Autonomous} {Driving}}}, ser. Signals and {Communication}
  {Technology}.\hskip 1em plus 0.5em minus 0.4em\relax Singapore: Springer
  Singapore, 2020. [Online]. Available:
  \url{http://link.springer.com/10.1007/978-981-13-9193-4}
\BIBentrySTDinterwordspacing

\end{thebibliography}
\bibliographystyle{IEEEtran}

\end{document}